\documentclass[apj]{emulateapj}
\usepackage{mathptmx}
\usepackage{lscape}
\usepackage{rotate}

\newcommand{\ltsim}{\raisebox{-.5ex}{$\;\stackrel{<}{\sim}\;$}}
\newcommand{\gtsim}{\raisebox{-.5ex}{$\;\stackrel{>}{\sim}\;$}}
\newcommand{\kms}{\ifmmode {\rm km\ s}^{-1} \else km s$^{-1}$\fi}
\newcommand{\lledd}{$L/L_{\rm Edd}$}

\newcommand{\et}{et al.\ }
\newcommand{\xray}{\hbox{X-ray}}
\newcommand{\aox}{$\alpha_{\rm ox}$}
\newcommand{\aro}{$\alpha_{\rm ro}$}
\newcommand{\daox}{$\Delta\alpha_{\rm ox}$}
\newcommand{\nh}{$N_{\rm H}$}

\newcommand{\xmm}{{\sl XMM-Newton}}
\newcommand{\chandra}{{\sl Chandra}}

\journalinfo{The Astrophysical Journal, 696:???--???, 2009 May 1}
\slugcomment{Received 2008 September 3; accepted 2009 January 28}

\shorttitle{X-RAY INSIGHTS INTO THE NATURE OF WLQS}
\shortauthors{SHEMMER ET AL.}

\begin{document}

\title{X-ray Insights into the Nature of Weak Emission-Line Quasars at
  High Redshift}

\author{
Ohad~Shemmer,\altaffilmark{1,2}
W.~N.~Brandt,\altaffilmark{1}
Scott~F.~Anderson,\altaffilmark{3}
Aleksandar~M.~Diamond-Stanic,\altaffilmark{4}
Xiaohui~Fan,\altaffilmark{4}
Gordon~T.~Richards,\altaffilmark{5}
Donald~P.~Schneider,\altaffilmark{1}
and Michael~A.~Strauss\altaffilmark{6}
}

\altaffiltext{1}
               {Department of Astronomy \& Astrophysics, The Pennsylvania State
               University, University Park, PA 16802}
\altaffiltext{2}
               {Current address: Department of Physics, University of
               North Texas, Denton, TX 76203; ohad@unt.edu}
\altaffiltext{3}
               {Department of Astronomy, University of Washington, Box 351580,
               Seattle, WA 98195}
\altaffiltext{4}
               {Steward Observatory, University of Arizona, 933 North Cherry
               Avenue, Tucson, AZ 85721}
\altaffiltext{5}
               {Department of Physics, Drexel University, 3141 Chestnut Street,
               Philadelphia, PA 19104}
\altaffiltext{6}
               {Princeton University Observatory, Peyton Hall, Princeton,
               NJ 08544}

\begin{abstract}
  We present \chandra\ observations of nine high-redshift quasars
  \hbox{($z=2.7-5.9$)} discovered by the Sloan Digital Sky Survey with
  weak or undetectable high-ionization emission lines in their UV
  spectra (WLQs). Adding archival \xray\ observations of six
  additional sources of this class has enabled us to place the
  strongest constraints yet on the \xray\ properties of this
  remarkable class of AGNs. Although our data cannot rule out the
  possibility that the emission lines are overwhelmed by a
  relativistically boosted continuum, as manifested by BL~Lac objects,
  we find that WLQs are considerably weaker in the \xray\ and radio
  bands than the majority of BL~Lacs found at much lower redshifts.
  If WLQs are high-redshift BL~Lacs, then it is difficult to explain
  the lack of a large parent population of \xray\ and radio bright
  weak-lined sources at high redshift. We also consider the
  possibility that WLQs are quasars with extreme properties, and in
  particular that the emission lines are suppressed by high accretion
  rates. Using joint spectral fitting of the \xray\ spectra of 11 WLQs
  we find that the mean photon index in the hard \xray\ band is
  consistent with those observed in typical radio-quiet AGNs with no
  hint of an unusually steep hard-\xray\ spectrum. This result poses a
  challenge to the hypothesis that WLQs have extremely high accretion
  rates, and we discuss additional observations required to test this
  idea.
\end{abstract}


\section{Introduction}
\label{introduction}

Since strong and broad emission lines are a hallmark of quasar
optical--UV spectra, the discovery of $\sim50$ Sloan Digital Sky
Survey (SDSS; York \et 2000) quasars at \hbox{$z\sim2.7-5.9$} with
extremely weak or undetectable emission lines (hereafter WLQs) is
puzzling (see Fan \et 1999 for the discovery of the prototype
high-redshift WLQ, SDSS~J153259.96$-$003944.1, at $z=4.62$). While the
rest-frame equivalent widths (EWs) of the Ly$\alpha +$\ion{N}{5}
emission lines of typical SDSS quasars at $z\sim3$ follow a
\hbox{log-normal} distribution and 68\% of EWs are in the range
$\sim40-100$~\AA\ (Fan \et 2001; \hbox{Diamond-Stanic} \et 2009), WLQs
exhibit EW(Ly$\alpha +$\ion{N}{5})$<10$~\AA, constituting a $4~\sigma$
deviation from the mean of the distribution (there is no detectable
corresponding deviation at the high EW end; \hbox{Diamond-Stanic} \et
2009). By virtue of their largely featureless spectra, the redshifts
of these WLQs can be determined only from the onset of the Ly$\alpha$
forest or the Lyman limit (i.e., $z>2.2$ for SDSS sources; e.g., Fan
\et 1999; Anderson \et 2001; Collinge \et 2005; Schneider \et 2005).

Since WLQs exhibit typical quasar UV continua with no signs of broad
\ion{C}{4} absorption, they are unlikely to be dust-obscured or
broad-absorption line quasars (see, e.g., Anderson \et 2001; Collinge
\et 2005). In addition, Shemmer \et (2006; hereafter S06) have argued
against the possibility that these sources are high-redshift galaxies
with apparent quasar-like luminosities due to gravitational-lensing
amplification, or lensed quasars with continua amplified by
microlensing. S06 suggested that the two most likely interpretations
for the weakness of the high-ionization emission lines in these
sources are either line dilution due to a relativistically boosted
continuum or, alternatively, physical suppression of the
high-ionization emission lines.

The first interpretation for the nature of WLQs involves relativistic
beaming of the continuum emission from a jet directed close to our
line of sight, as observed in BL~Lacertae objects (BL~Lacs). This
scenario implies that the SDSS may have discovered the long-sought,
high-redshift BL~Lacs (e.g., Stocke 2001); the absence of such sources
is even more puzzling since their beamed continua should have made
them easier to detect at high redshift than ordinary quasars. However,
this interpretation faces serious difficulties. For example, S06 have
found that the basic \xray, optical, and radio properties of five WLQs
are consistent only with the \xray- and radio-weak tail of the BL~Lac
population. Moreover, while both WLQs and BL~Lacs share the property
of a lineless UV spectrum, BL~Lacs also often exhibit high optical
polarization levels, rapid and large-amplitude flux variations, and
spectral energy distributions (SEDs) dominated by beamed, non-thermal
emission at all wavelengths (e.g., Urry \& Padovani 1995). Such
extreme properties have, so far, not been observed in a subsample of
SDSS WLQs (e.g., Fan \et 1999; Diamond-Stanic \et 2009).

\begin{figure*}
\epsscale{1.2}
 \plotone{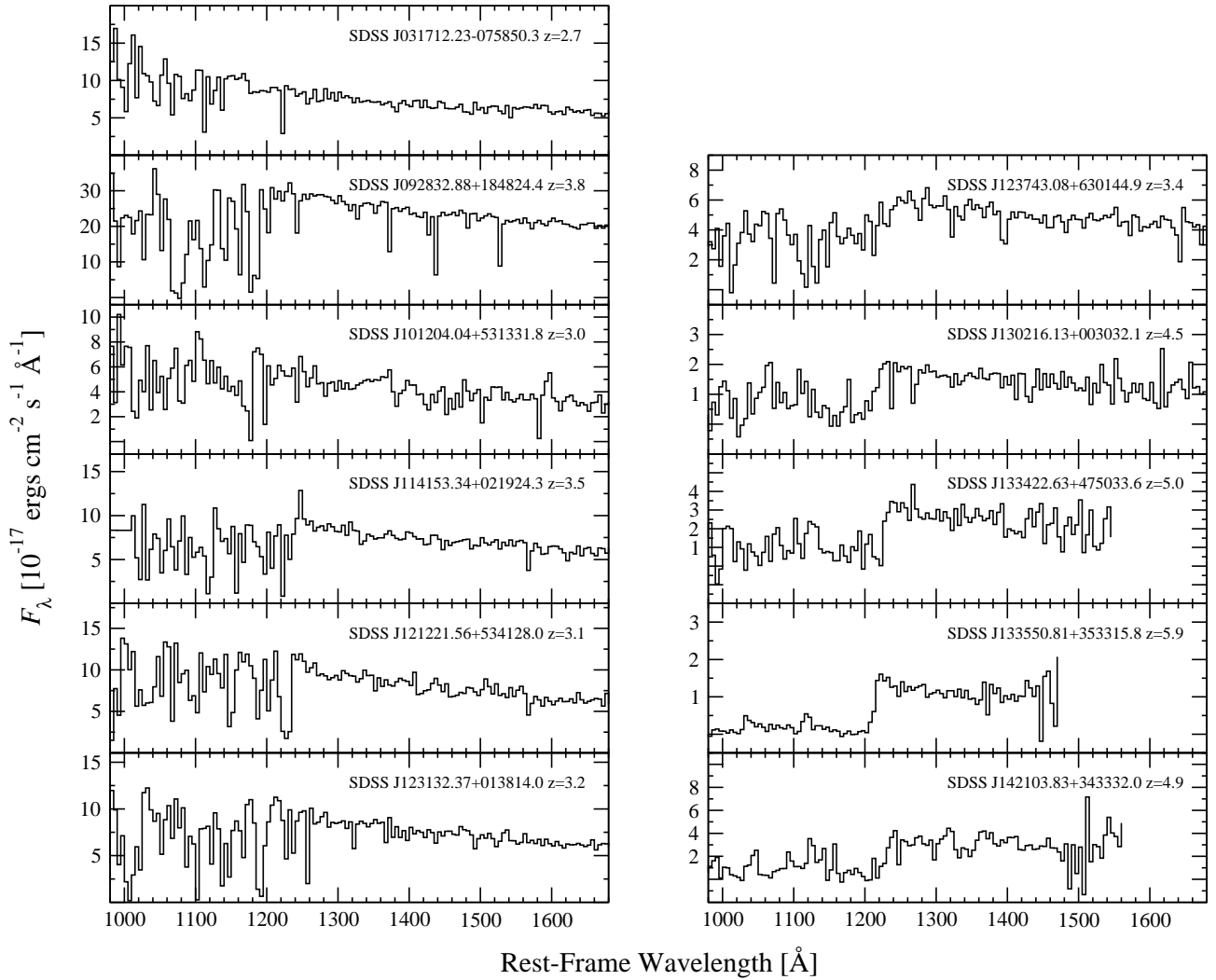}
 \caption{Ultraviolet spectra of the WLQ sample resampled in bins of
   5\,\AA, for clarity. Spectra were obtained from the SDSS archive,
   except for SDSS~J1335$+$3533, for which the spectrum was obtained
   from the Keck observatory (Fan \et 2006).}
\label{spectra}
\end{figure*}

The alternative interpretation is that WLQs may be quasars with
abnormal photoionizing continuum or broad emission-line region (BELR)
properties. In this scenario, the weak high-ionization emission lines
of WLQs may, for example, ultimately be a consequence of exceptionally
high accretion rates in these sources. A higher accretion rate is
expected to result in a softer, UV-peaked continuum that suppresses
the high-ionization emission lines, while low-ionization species such
as the Balmer and \ion{Fe}{2} lines are not significantly affected
(e.g., Leighly \et 2007a,b). Accretion rates in WLQs have not yet been
determined in order to test this possibility (for example, by
measuring the width of H$\beta$ in such sources; see, e.g., Shemmer
\et 2004).

The key to understanding WLQs, therefore, lies in comparing their
multiwavelength spectral, as well as polarimetric and temporal,
properties with those of BL~Lacs and typical quasars. While the SDSS
WLQs are rare, their exceptional properties constitute a challenge to
our overall understanding of AGN physics and thus might lead to new
insights about the accretion process, emission-line formation, and
relativistic jets. In this paper, we measure basic \xray\ properties
for the first respectably sized sample of WLQs in order to shed light
on their nature and to assist in planning more detailed \xray\
spectroscopy of these sources in the future. Since \hbox{X-rays} probe
the innermost regions of AGNs, our observations are intended to study
the basic diagnostics of the central accretion process in WLQs. This
paper nearly triples the number of $z>2.2$ WLQs that have sensitive
\xray\ coverage, increasing the number of \xray\ detections by almost
a factor of five.

This paper is organized as follows: in \S~\ref{observations} we
describe the WLQ sample selection, as well as our \xray\ observations
and their reduction; our basic findings are presented in
\S~\ref{results}. In \S~\ref{discussion} we discuss the main results
emerging from our \xray\ study of WLQs and the clues they provide on
the nature of these sources; a summary is given in \S~\ref{summary}.
Complete WLQ names~are given in the tables and figures, and their
abbreviated versions are used throughout the text.  Throughout this
work we compute luminosity distances using the standard cosmological
model with parameters $H_{0}=70$~\kms~Mpc$^{-1}$,
$\Omega_{\Lambda}=0.7$, and $\Omega_{M}=0.3$.

\begin{deluxetable*}{lcclccc}
\tablecolumns{7} 
\tablewidth{0pc}
\tablecaption{\chandra\ Observation Log}
\tablehead{
\colhead{Object} &
\colhead{} &
\colhead{$\Delta_{\rm Opt-X}$\tablenotemark{a}} & 
\colhead{\xray} &
\colhead{\chandra} &
\colhead{Exp.~Time\tablenotemark{b}} &
\colhead{} \\
\colhead{(SDSS~J)} &
\colhead{$z$} &
\colhead{(arcsec)} &
\colhead{Obs. Date} & 
\colhead{Obs. ID} &
\colhead{(ks)} &
\colhead{Ref.}
}
\startdata
031712.23$-$075850.3 & 2.7 & 0.1 & 2007 Mar 17 & 7780 & 4.42 & 1 \\
092832.90$+$184824.4 & 3.8 & 0.1 & 2007 Jan 2 & 7779 & 3.98 & 2 \\
114153.34$+$021924.3 & 3.5 & 0.2 & 2007 Feb 10 & 7777 & 3.99 & 1 \\
121221.56$+$534128.0 & 3.1 & 0.4 & 2007 Jul 24 & 7778 & 4.13 & 1 \\
123132.38$+$013814.0 & 3.2 & 0.1 & 2007 Jul 5 & 7781 & 4.09 & 1 \\
130216.13$+$003032.1\tablenotemark{c} & 4.5 & {\phn}0.6\tablenotemark{d} &
2007 Feb 24 & 7784 & 10.50 & 3 \\
133422.63$+$475033.6 & 5.0 & 0.1 & 2007 Aug 25 & 7885 & 11.60 & 3 \\
133550.81$+$353315.8 & 5.9 & 0.2 & 2008 Mar 10 & 7783 & 23.47 & 4 \\
142103.83$+$343332.0 & 4.9 & 0.3 & 2008 Feb 27 & 7782 & 12.79 & 5
\enddata
\tablecomments{The optical positions of the quasars have been obtained
from the reference given in the seventh column, and the \xray\
positions have been obtained with {\sc wavdetect}.}
\tablenotetext{a}{Angular distance between the optical and \xray\
  positions.}
\tablenotetext{b}{The \chandra\ exposure time has been corrected for
  detector dead time.}
\tablenotetext{c}{A 10.81~ks \chandra\ Cycle~4 exposure of the source
  (Obs. ID~3958) is not reflected in this Table. The Cycle~8
  observation raises the total exposure time on the source to
  21.31~ks, and the merged event file is used in subsequent analyses
  throughout the paper (see S06 and \S~\ref{observations} for
  details).}
\tablenotetext{d}{Based on a merged image composed of \chandra\
  Cycle~4 and Cycle~8 exposures.}
\tablerefs{(1) Collinge \et (2005); (2) this work; (3) Schneider \et
  (2005); (4) Fan \et (2006); (5) Schneider \et (2007).}
\label{log}
\end{deluxetable*}

\section{Sample Selection, Observations, and Data Reduction}
\label{observations}

\subsection{Targeted \chandra\ observations}
\label{chandra_obs}

We selected nine WLQs at $z=2.7-5.9$ for short \hbox{($\sim4-23$~ks)}
\chandra\ observations in Cycle~8 (2007--2008); the rest-frame
ultraviolet spectra of these sources appear in Fig.~\ref{spectra}.
The \chandra\ observation log appears in Table~\ref{log}, where we
also list the papers in which the sources were first identified as
WLQs or BL~Lac candidates. These sources were selected from the SDSS
archive for being the brightest sources of this class; except for
SDSS~J1302$+$0030, these were not targeted in \hbox{X-rays} prior to
Cycle~8. SDSS~J1302$+$0030 was not detected in a 10.81~ks \chandra\
Cycle~4 exposure; our additional 10.50~ks exposure in Cycle~8 has
enabled its detection (see below).  SDSS~J1335$+$3533 was discovered
by Fan \et (2006) and is the only source in this sample that was not
detected with the SDSS quasar-selection algorithm (Richards \et 2002;
this algorithm is not designed to detect quasars above $z\simeq5.4$).

The nine WLQs were observed with the Advanced CCD Imaging Spectrometer
(ACIS; Garmire \et 2003) with the S3 CCD at the aimpoint. Faint mode
was used for the event telemetry format in all the observations, and
{\sl ASCA} grade 0,~2,~3,~4,~and~6 events were used in the analysis,
which was carried out using standard {\sc ciao\footnote{\chandra\
    Interactive Analysis of
    Observations. \\ http://cxc.harvard.edu/ciao/} v3.4} routines. No
background flares are present in these observations. Source detection
was carried out with {\sc wavdetect} (Freeman \et 2002) using wavelet
transforms (with wavelet scale sizes of 1,~1.4,~2,~2.8,~and~4 pixels)
and a false-positive probability threshold of 10$^{-4}$. Given the
small number of pixels being searched due to the accurate a priori
source positions and the subarcsecond on-axis angular resolution of
\chandra, the probability of spurious detections is extremely low;
most of the sources were in fact detected at a false-positive
probability threshold of 10$^{-6}$. The \xray\ positions of the
detected sources differ by \hbox{0.1\arcsec--0.6\arcsec} from their
optical positions (Table~\ref{log}). These results are consistent with
expected \chandra\ positional errors.

The \xray\ counts detected in the ultrasoft band
(\hbox{0.3--0.5~keV}), soft band (\hbox{0.5--2~keV}), hard band
(\hbox{2--8~keV}), and full band (\hbox{0.5--8~keV}) are reported in
Table~\ref{counts}. The counts were derived from manual aperture
photometry, and they are consistent with the photometry obtained using
{\sc wavdetect}. Except for SDSS~J1302$+$0030, all of the targeted
WLQs were detected with 3--110 full-band counts. A significant
detection of SDSS~J1302$+$0030 (using {\sc wavdetect} with a
false-positive probability threshold of 10$^{-4}$) was achieved only
by merging the Cycle~8 exposure (with 3 full-band counts) with that
from Cycle~4 (with 2 full-band counts; see S06); the \xray\ counts for
this source, presented in Table~\ref{counts}, were obtained from the
merged image. Table~\ref{counts} also includes the band ratio, which
is the ratio between the counts in the hard band and the counts in the
soft band, and the effective power-law photon index [$\Gamma$, where
$N(E)\propto E^{-\Gamma}$; $\Gamma$ was calculated from the band ratio
using the \chandra\ {\sc pimms} v3.9d
tool\footnote{http://cxc.harvard.edu/toolkit/pimms.jsp}] for each
source.

Finally, we searched for rapid variability within the \chandra\
observations of all WLQs detected with $>10$ full-band counts by
applying a Kolmogorov-Smirnov (KS) test to the photon arrival times
against the null hypothesis of a constant rate, but no significant
flux variations were detected. This result is not unexpected given the
combination of the relatively short observations (15--30~min in the
rest-frame) and the small number of detected photons from most of the
sources.

\begin{deluxetable*}{lcccccc}
\tablecolumns{7}
\tablewidth{0pt}
\tablecaption{X-ray Counts, Band Ratios, and Effective Photon Indices}
\tablehead{ 
\colhead{Object} &
\multicolumn{4}{c}{X-ray Counts$^{\rm a}$} \\
\cline{2-5} \\
\colhead{(SDSS~J)} &
\colhead{0.3--0.5~keV} &
\colhead{0.5--2~keV} & 
\colhead{2--8~keV} &
\colhead{0.5--8~keV} & 
\colhead{Band Ratio\tablenotemark{b}} &
\colhead{$\Gamma$\tablenotemark{b}}
}
\startdata
031712.23$-$075850.3 & $<3.0$ & {\phn}23.9$^{+6.0}_{-4.8}$ &
{\phn}3.0$^{+2.9}_{-1.6}$ & {\phn}26.8$^{+6.3}_{-5.1}$ &
{\phn}0.13$^{+0.13}_{-0.07}$ & {\phn}2.6$^{+0.8}_{-0.6}$ \\
092832.90$+$184824.4 & $<3.0$ & {\phn}52.2$^{+8.3}_{-7.2}$ &
{\phn}10.9$^{+4.4}_{-3.2}$ & {\phn}63.1$^{+9.0}_{-7.9}$ &
0.21$^{+0.09}_{-0.07}$ & {\phn}2.2$^{+0.4}_{-0.3}$ \\
114153.34$+$021924.3 & $<4.8$ &{\phn}28.6$^{+6.4}_{-5.3}$ &
{\phn}12.7$^{+4.7}_{-3.5}$ & {\phn}41.4$^{+7.5}_{-6.4}$ &
{\phn}0.44$^{+0.19}_{-0.15}$ & {\phn}1.4$^{+0.4}_{-0.3}$ \\
121221.56$+$534128.0 & $<3.0$ & {\phn}3.0$^{+2.9}_{-1.6}$ & $<3.0$ &
{\phn}3.0$^{+2.9}_{-1.6}$ & $<1.00$ & $>0.7$ \\
123132.38$+$013814.0 & {\phn}10.9$^{+4.4}_{-3.2}$ &
{\phn}83.6$^{+10.2}_{-9.1}$ & {\phn}26.5$^{+6.2}_{-5.1}$ &
{\phn}109.9$^{+11.5}_{-10.4}$ & {\phn}0.32$^{+0.08}_{-0.07}$ &
{\phn}1.7$^{+0.2}_{-0.2}$ \\
130216.13$+$003032.1\tablenotemark{c} & $<3.0$ &
{\phn}2.0$^{+2.7}_{-1.3}$ & $<8.0$ & {\phn}5.0$^{+3.4}_{-2.2}$ &
$<4.00$ & $>-0.6$ \\
133422.63$+$475033.6 & $<4.8$ & {\phn}13.8$^{+4.8}_{-3.7}$ &
{\phn}4.0$^{+3.2}_{-1.9}$ & {\phn}17.7$^{+5.3}_{-4.2}$ &
0.29$^{+0.25}_{-0.16}$ & {\phn}1.8$^{+0.7}_{-0.6}$ \\
133550.81$+$353315.8 & {\phn}$<3.0$ &{\phn}3.0$^{+2.9}_{-1.6}$ &
{\phn}$<6.4$ & {\phn}5.0$^{+3.4}_{-2.2}$ & {\phn}$<2.13$ &
{\phn}$>0.0$ \\
142103.83$+$343332.0 & {\phn}$<3.0$ & {\phn}2.0$^{+2.7}_{-1.3}$ &
{\phn}$<4.8$ & {\phn}3.8$^{+3.1}_{-1.9}$ &
{\phn}$<2.40$ & {\phn}$>-0.1$
\enddata
\tablenotetext{a}{Errors on the \xray\ counts were computed according
  to Tables~1 and 2 of Gehrels (1986) and correspond to the 1$\sigma$
  level; these were calculated using Poisson statistics.  The upper
  limits are at the 95\% confidence level and were computed according
  to Kraft, Burrows, \& Nousek (1991). Upper limits of 3.0, 4.8, 6.4,
  and 8.0 indicate that 0, 1, 2, and 3 \xray\ counts, respectively,
  have been found within an extraction region of radius 1\arcsec\
  centered on the optical position of the quasar (considering the
  background within this source-extraction region to be negligible).}
\tablenotetext{b}{We calculated errors at the $1~\sigma$ level for the
  band ratio (the ratio between the \hbox{2--8~keV} and
  \hbox{0.5--2~keV} counts) and effective photon index following the
  ``numerical method'' described in $\S$~1.7.3 of Lyons (1991); this
  avoids the failure of the standard approximate-variance formula when
  the number of counts is small (see $\S$~2.4.5 of Eadie et al. 1971).
  The photon indices have been obtained applying the correction
  required to account for the ACIS quantum-efficiency decay at low
  energy.}
\tablenotetext{c}{The \xray\ counts, band ratio, and photon index for
  this source were obtained from a merged event file of our new
  \chandra\ Cycle~8 observation and an archival \chandra\ Cycle~4
  observation (see \S~\ref{observations} for more details). The total
  exposure time of the merged event file is 21.31~ks.}
\label{counts}
\end{deluxetable*}

\subsection{Serendipitous \xmm\ observations}
\label{xmm_obs}

Two additional SDSS WLQs at $z>2.2$ were serendipitously observed by
\xmm\ (Jansen \et 2001). SDSS~J1012$+$5313 at $z=3.0$ is a newly
identified WLQ from SDSS Data Release~5 (Schneider \et 2007). The
source was observed with \xmm\ on 2001 April 19 for 18.7~ks (dataset
ID 0111100201; PI M.~Watson). SDSS~J1237$+$6301 at $z=3.4$ was
identified as a BL~Lac candidate by Collinge \et (2005). The source
was observed with \xmm\ on 2006 October 25 for 20.7~ks (dataset ID
0402250101; PI B.~Maughan). The \xmm\ observations were processed
using standard \xmm\ Science Analysis
System\footnote{http://xmm.esac.esa.int/sas} v7.1.0 tasks. The
SDSS~J1012$+$5313 observation included only MOS1 and MOS2 images, and
the entire observation was carried out during a period of flaring
activity; hence no time filtering was employed. The SDSS~J1237$+$6301
observation was filtered in time to remove periods of background
flaring in which the count rates of each MOS (pn) detector exceeded
0.35 (1.0) counts~s$^{-1}$ for events having $E>10$\,keV; the net
exposure times were 10.7~ks and 4.9~ks for the MOS1/MOS2 and pn
detectors, respectively. We detected SDSS~J1012$+$5313 at 7.0\arcmin\
from the \xmm\ aimpoint in both the MOS1 and MOS2 detectors. For each
detector, source counts were extracted from an aperture with a
30\arcsec\ radius centered on the source, and background counts were
extracted from an aperture with a similar size in a nearby source-free
region. SDSS~J1237$+$6301 was not detected in any of the \xmm\
detectors. The upper limit on the \xray\ flux of the source was taken
as the square root of the number of counts encircled within an
aperture of 30\arcsec\ centered on the optical position of the source
(which is offset by 9.8\arcmin\ from the aimpoint) multiplied by three
(i.e., a $3 \sigma$ upper limit). The \xray\ counts of
SDSS~J1012$+$5313 as well as the upper limit on the number of counts
of SDSS~J1237$+$6301 were corrected for vignetting using the exposure
maps due to the large offsets from the aimpoints. The SDSS spectra of
the two sources appear in Figure~\ref{spectra}.

\section{Results}
\label{results}

\subsection{Basic Multiwavelength Properties of the WLQ Sample}
\label{basic}

Table~\ref{properties} (placed at the end of the paper) lists the
basic \xray, optical, and radio properties of our WLQ sample: \\
{\sl Column (1)}. --- The SDSS~J2000.0 quasar coordinates, accurate to
$\sim0.1$\arcsec\ (Pier \et 2003). \\
{\sl Column (2)}. --- The Galactic column density in units of
\hbox{$10^{20}$~cm$^{-2}$}, taken from Dickey \& Lockman (1990) and
obtained with the HEASARC \nh\
tool.\footnote{http://heasarc.gsfc.nasa.gov/cgi-bin/Tools/w3nh/w3nh.pl}
\\
{\sl Column (3)}. --- The monochromatic $AB$ magnitude at a rest-frame
wavelength of 1450\,\AA\
(\hbox{$AB_{1450}=-2.5\log~f_{1450~\mbox{\scriptsize\AA}}-48.6$}; Oke
\& Gunn 1983). All the magnitudes have been corrected for Galactic
extinction (Schlegel, Finkbeiner, \& Davis 1998). Except for
SDSS~J1335$+$3533, for which this magnitude has been obtained from a
Keck spectrum, all the magnitudes were obtained from SDSS Data
Release~5 spectra and a fiber light-loss correction has been applied;
this was calculated as the difference between the synthetic $i$
magnitude (i.e., the integrated flux across the $i$ bandpass in the
SDSS spectrum) and the photometric SDSS $i$ magnitude, assuming no
flux variation between the photometric and spectroscopic epochs. \\
{\sl Column (4)}. --- The absolute $B$-band magnitude, computed
assuming a \hbox{UV--optical} \hbox{($\sim1200-4000$~\AA)} power-law
slope of $\alpha_{\rm UV}=-0.5$ \hbox{($f_{\nu}\propto
  \nu^{\alpha_{\rm UV}}$}; Vanden
Berk \et 2001). \\
{\sl Columns (5) and (6)}. --- The flux density and monochromatic
luminosity at a rest-frame wavelength of 2500~\AA, computed from the
magnitudes in Col.~3, assuming a \hbox{UV--optical} power-law slope as
in Col.~4. \\
{\sl Columns (7) and (8)}. --- The count rate in the observed-frame
\hbox{0.5--2~keV} band and the corresponding flux, corrected for
Galactic absorption.  The fluxes have been calculated using {\sc
  pimms}, assuming a power-law model with $\Gamma=2.0$. \\
{\sl Columns (9) and (10)}. --- The flux density and monochromatic
luminosity at a rest-frame energy of 2~keV, computed assuming
$\Gamma=2.0$ in the entire \chandra\ band ($0.5-8$~keV). \\
{\sl Column (11)}. --- The luminosity in the rest-frame
\hbox{2--10~keV} band. \\
{\sl Column (12)}. --- The optical-to-\xray\ power-law slope, \aox,
defined as:

\begin{equation}
\alpha_{\rm ox}=\frac{\log(f_{\rm
    2~keV}/f_{2500\mbox{\rm~\scriptsize\AA}})} {\log(\nu_{\rm
    2~keV}/\nu_{2500\mbox{\rm~\scriptsize\AA}})}
\label{eq:aox}
\end{equation}
where $f_{\rm 2~keV}$ and $f_{2500\mbox{\rm~\scriptsize\AA}}$ are the
flux densities at rest-frame 2~keV and 2500~\AA, respectively. \\
{\sl Column (13)}. --- The difference between the measured \aox\ (from
Col.~12) and the predicted \aox, given the UV luminosity from Col.~6,
based on the observed \hbox{aox-$L_{\nu}(2500~\mbox{\AA})$} relation
in AGNs (given as Eq.~3 of Just \et 2007; see also Steffen \et
2006). The statistical significance of this difference is also given
in units of $\sigma$, where \hbox{$\sigma=0.146$} for \hbox{$31<\log
  L_{\nu}(2500~\mbox{\AA})<32$}, and \hbox{$\sigma=0.131$} for
\hbox{$32<\log L_{\nu}(2500~\mbox{\AA})<33$} (see Table~5 of Steffen
\et 2006; these differences do not account for
flux-density measurement~errors). \\
{\sl Column (14)}. --- The radio-to-optical power-law slope, \aro,
defined as:

\begin{equation}
\alpha_{\rm ro}=\frac{\log(f_{\rm
    5~GHz}/f_{2500\mbox{\rm~\scriptsize\AA}})} {\log(\nu_{\rm
    5~GHz}/\nu_{2500\mbox{\rm~\scriptsize\AA}})}
\label{eq:aro}
\end{equation}
This parameter can be directly converted to the radio-loudness
parameter $R$ (where \hbox{$R=f_{\rm 5~GHz}/f_{\rm
    4400~\mbox{\scriptsize\AA}}$}; Kellermann \et 1989) in the
following way:
\begin{equation}
  \alpha_{\rm ro}=\frac{\log \left [ R \left ( \frac{2500}{4400} \right )
^{\alpha_{\rm UV}} \right ] }
  {\log(\nu_{\rm 5~GHz}/\nu_{2500\mbox{\rm~\scriptsize\AA}})}
\label{eq:aroR}
\end{equation}
where we take $\alpha_{\rm UV}=-0.5$ (as in Col.~4). The flux density
at a rest-frame frequency of 5~GHz was computed from the FIRST
(Becker, White, \& Helfand 1995) flux density at an observed-frame
frequency of 1.4~GHz, if available. Otherwise, the NVSS (Condon et
al. 1998) flux density at the same observed band was used. We assume a
radio power-law slope of $\alpha_{\rm r}=-0.5$ \hbox{($f_{\nu}\propto
  \nu^{\alpha_{\rm r}}$}; e.g., Rector \et 2000). Upper limits on
\aro\ are at the $3~\sigma$ level, since the positions of all our
sources are known a priori. The flux densities at rest frame 4400~\AA\
were computed from the $AB_{1450}$ magnitudes assuming $\alpha_{\rm
  UV}=-0.5$ (as in Col.~4). Typical \aro\ values are $<-0.39$ (i.e.,
$R>100$) for radio-loud quasars (RLQs) and $>-0.21$ (i.e., $R<10$) for
radio-quiet quasars (RQQs). Seven of our sources are RQQs; four
sources are radio intermediate with $-0.39 \ltsim $\aro$ \ltsim -0.21$
(i.e., $10\ltsim R \ltsim 100$).

\begin{figure}
\includegraphics[width=3.8cm,angle=-90]{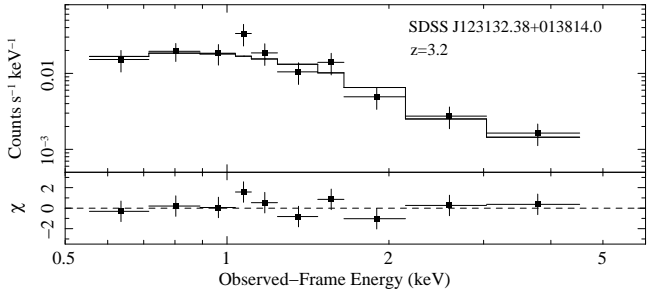}
 \caption{\chandra\ spectrum of SDSS\,J1231$+$0138. Filled squares
   represent the binned spectrum, and the solid histogram represents
   the best-fit Galactic-absorbed power-law model. The $\chi$
   residuals are in units of $\sigma$ with error bars of size 1.}
\label{sdssj1231_spec}
\end{figure}

\setcounter{table}{3}

\subsection{The X-ray Spectrum of SDSS~J1231$+$0138}
\label{sdssj1231}

SDSS~J1231$+$0138 is the only WLQ in our sample with sufficient \xray\
counts to allow an investigation of its \xray\ spectrum. The spectrum
of the source was extracted with the {\sc ciao} task {\sc psextract}
using a circular region of 2\arcsec\ radius centered on the \xray\
centroid. The background was extracted from a source-free annulus with
inner and outer radii of 5\arcsec\ and 25\arcsec, respectively. The
spectrum was binned in groups of 10 counts per bin. We used {\sc xspec
  v11.3.2} (Arnaud 1996) to fit the spectrum with a power-law model
and a Galactic-absorption component, which was kept fixed during the
fit (\nh$=1.81 \times 10^{20}$\,cm$^{-2}$; Dickey \& Lockman
1990). The spectrum and its best-fit model and residuals appear in
Fig.\,\ref{sdssj1231_spec}. The best-fit photon index is
$\Gamma=1.76^{+0.35}_{-0.34}$ with $\chi^2=5.57$ for 8 degrees of
freedom; this photon index is consistent with the value obtained from
the band ratio (Table~\ref{counts}) and is typical for
radio-intermediate AGNs such as SDSS~J1231$+$0138 (\aro$=-0.32$; see
\S\,\ref{basic}; e.g., Reeves \& Turner 2000). Adding an
intrinsic-absorption component at the redshift of the source did not
improve the fit; the upper limit on the intrinsic absorption is \nh
\ltsim5.70$\times10^{22}$\,cm$^{-2}$ (at 90\% confidence).

\begin{figure}
\includegraphics[width=7.5cm,angle=-90]{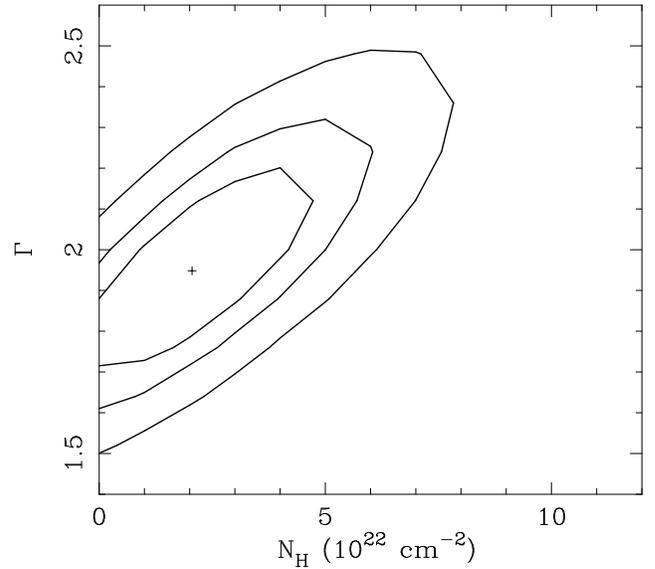}
\caption{68\%, 90\%, and 99\% confidence regions for the photon index
  vs. intrinsic column density derived from joint spectral fitting of
  our sample of 11 WLQs over their entire observed energy range.}
\label{Gamma_NH}
\end{figure}

\subsection{Mean X-ray Spectral Properties of High-Redshift WLQs}
\label{sec_joint_fit}

Since all of our WLQs, except for SDSS~J1231$+$0138 (see
\S\,\ref{sdssj1231}), lack sufficient \xray\ counts for individual
spectral analyses, we have obtained mean \xray\ spectral constraints
for high-redshift WLQs as a group by jointly fitting their \xray\
spectra (e.g., Vignali \et 2005; S06). Eleven sources with \chandra\
spectra were included in the joint-fitting process, including the nine
WLQs targeted in this work (i.e., all sources from Tables~\ref{log}
and~\ref{counts} including the merged dataset of SDSS~J1302$+$0030)
and the only other two \chandra-detected WLQs from S06, namely
SDSS~J140850.91$+$020522.7 and SDSS~J144231.72$+$011055.2 (see
\hbox{Tables~1--3} of S06). The combined spectrum consists of a total
of 350 full-band counts, obtained in 107~ks. We note that there are
only two other \xray-detected WLQs with \xmm\ spectra:
SDSS~J1012$+$5313 (see \S\,\ref{xmm_obs}) and
SDSS~J004054.65$-$091526.8 (S06); these spectra do not provide a
significant contribution to the spectral constraints since the first
was taken during a period of flaring activity (\S\,\ref{xmm_obs}) and
the second was detected at low signal-to-noise (with only
31.8$\pm$11.6 net source counts; Schneider \et 2003).

\begin{deluxetable*}{llccc}
\tablecolumns{5} 
\tablewidth{0pt}
\tablecaption{Best-Fit Parameters from Joint Fitting of High-Redshift
  WLQs}
\tablehead
{
\colhead{} &
\colhead{} &
\colhead{Intrinsic \nh} &
\colhead{} &
\colhead{} \\
\colhead{{\sc xspec} Model} &
\colhead{Sources Included} &
\colhead{(10$^{22}$\,cm$^{-2}$)} &
\colhead{$\Gamma$} &
\colhead{$C$-statistic (bins)}
}
\startdata
{\sc phabs+pow} & all (11 sources) & \nodata & $1.79^{+0.17}_{-0.16}$
& 191.4 (294) \\
{\sc phabs+pow}, common energy range\tablenotemark{a} & all (11
sources) & \nodata & $1.96^{+0.22}_{-0.22}$ & 170.4 (260) \\
{\sc phabs+pow} & all but SDSS~J1231$+$0138 & \nodata &
$1.79^{+0.22}_{-0.22}$ & 133.0 (214) \\
{\sc phabs+pow} & seven radio-quiet WLQs\tablenotemark{b} & \nodata &
$1.81^{+0.45}_{-0.43}$ & 48.2 (84) \\
{\sc phabs+zphabs+pow} & all (11 sources) & $\le5.08$ &
$1.95^{+0.29}_{-0.28}$ & 189.9 (294) \\
{\sc phabs+zphabs+pow}, common energy range\tablenotemark{a} & all (11
sources) & $\le4.34$ & $1.96^{+0.26}_{-0.22}$ & 170.4 (260) \\
{\sc phabs+zphabs+pow} & all but SDSS~J1231$+$0138 & $\le7.05$ &
$1.97^{+0.37}_{-0.35}$ & 131.8 (214) \\
{\sc phabs+zphabs+pow} & seven radio-quiet WLQs\tablenotemark{b} &
$\le7.74$ & $1.86^{+0.72}_{-0.48}$ & 48.2 (84)
\enddata
\tablenotetext{a}{The fitting was performed only for the common
  rest-frame energy range among all sources
  (\hbox{3.45--29.60\,keV}).}
\tablenotetext{b}{Excluding the radio-intermediate WLQs
  SDSS~J0928$+$1848, SDSS~J1141$+$0219, SDSS~J1231$+$0138, and
  SDSS~J144231.72$+$011055.2.}
\label{joint_fit_par}
\end{deluxetable*}

The \xray\ spectra of all 11 WLQs were extracted as
in~\S\,\ref{sdssj1231} with {\sc psextract} using circular regions of
2\arcsec\ radius centered on the \xray\ centroid of each source, and
background counts were extracted using annuli of different sizes (to
avoid contamination from nearby \xray\ sources) centered on each
source. We used {\sc xspec} to fit the set of 11 unbinned spectra
jointly with the following models: (i) a power-law model and a
Galactic-absorption component (Dickey \& Lockman 1990; using the {\sc
  phabs} absorption model in {\sc xspec}), which was kept fixed during
the fit, and (ii) a model similar to the first with an added intrinsic
(redshifted) neutral-absorption component (using the {\sc zphabs}
model in {\sc xspec}). We used the \hbox{$C$-statistic} (Cash 1979) in
all the fits. The joint-fitting process associated with each quasar
its own Galactic-absorption column density and redshift, and allowed
the flux normalizations to vary freely while maintaining a fixed value
of the power-law photon index. The errors associated with the best-fit
\xray\ spectral parameters are quoted at the 90\% confidence level~for
one parameter of interest ($\Delta C=2.71$; Avni 1976; Cash 1979).

The joint-fitting process was carried out four times for both models.
In the first run, we included the entire observed-energy range of all
the spectra (\hbox{0.5--8~keV}). In the second run, we included only
the common rest-frame energy range of all 11 sources: 3.45\ltsim
$E_{\rm rest}$\ltsim29.60~keV (this range is based upon the entire
\hbox{0.5--8~keV} observed-frame band for each source, and is a
consequence of the wide range of redshifts among the sources; see also
S06). In the third run, we fitted the entire observed-energy range of
ten WLQs, excluding SDSS~J1231$+$0138 (since this source contributes
$\sim30$\% of the total counts and hence might bias the results). In
the fourth run, we have restricted the fitting only to radio-quiet
WLQs (i.e., excluding SDSS~J0928$+$1848, SDSS~J1141$+$0219,
SDSS~J1231$+$0138, and SDSS~J144231.72$+$011055.2) to minimize
potential contribution from jet-related emission.

Table~\ref{joint_fit_par} lists the best-fit parameters from the
joint-fitting process, and a contour plot of the $\Gamma$-\nh\
parameter space for the first run is shown in Figure~\ref{Gamma_NH}.
Based on $F$-tests, we find that the inclusion of an
intrinsic-absorption component (in addition to a Galactic absorbed
power-law model) does not significantly improve any of the fits; hence
no significant intrinsic absorption is detected in any of the joint
fits. Table~\ref{joint_fit_par} shows the mean photon index of WLQs is
not significantly different from typical values for RQQs or
radio-intermediate AGNs (e.g., Reeves \& Turner 2000; Page \et 2005;
Vignali \et 2005). In particular, there is no hint of an exceptionally
high photon index as might have been expected if WLQs are high
accretion-rate sources (see \S\,\ref{extreme}). The hard-\xray\
spectra of AGN can appear flattened if Compton reflection is not taken
into account in the spectral fitting. Since the fractional
contribution from Compton reflection generally weakens with increasing
luminosity, we do not expect that the typical value we obtain for the
mean photon index of our luminous, high-redshift WLQs is strongly
affected by Compton reflection (e.g., Iwasawa \& Taniguchi 1993; Zhou
\& Wang 2005; Bianchi \et 2007; Shemmer \et 2008).

The exclusion of SDSS~J1231$+$0138 or the exclusion of all
radio-intermediate sources have no significant effect on the results
(see also \S\,\ref{sdssj1231}). In addition, photon indices derived in
individual runs are consistent with each other, within the errors;
these errors are large due to the limited number of sources in our
sample and the short exposures. We conclude that the mean \xray\
spectral properties of high-redshift WLQs are consistent with those
observed in typical, radio- quiet-to-intermediate AGNs.

\section{Discussion}
\label{discussion}

Our primary goal is to use \xray\ information to shed light on the
nature of WLQs at high redshift. The \chandra\ and serendipitous \xmm\
observations have provided significantly stronger constraints on the
\xray\ luminosities of our WLQs than previous {\sl ROSAT} observations
(see also Collinge \et 2005; S06). This should allow us to scrutinize
the classification of the WLQs and determine the most probable
explanation for the absence of their high-ionization lines. The two
most likely scenarios for the absence of the UV lines are the
association of WLQs with BL~Lac objects or the hypothesis that WLQs
are quasars with extreme properties (see \S\,\ref{introduction}; S06).
Given the \xray, optical, and radio luminosities presented in this
work for an extended sample of 15 high-redshift WLQs (i.e., including
11 sources from this work and four sources from S06), we are now in a
position to perform meaningful comparisons of these basic properties
with those of BL~Lacs.

\begin{figure*}
\epsscale{1.0}
 \plotone{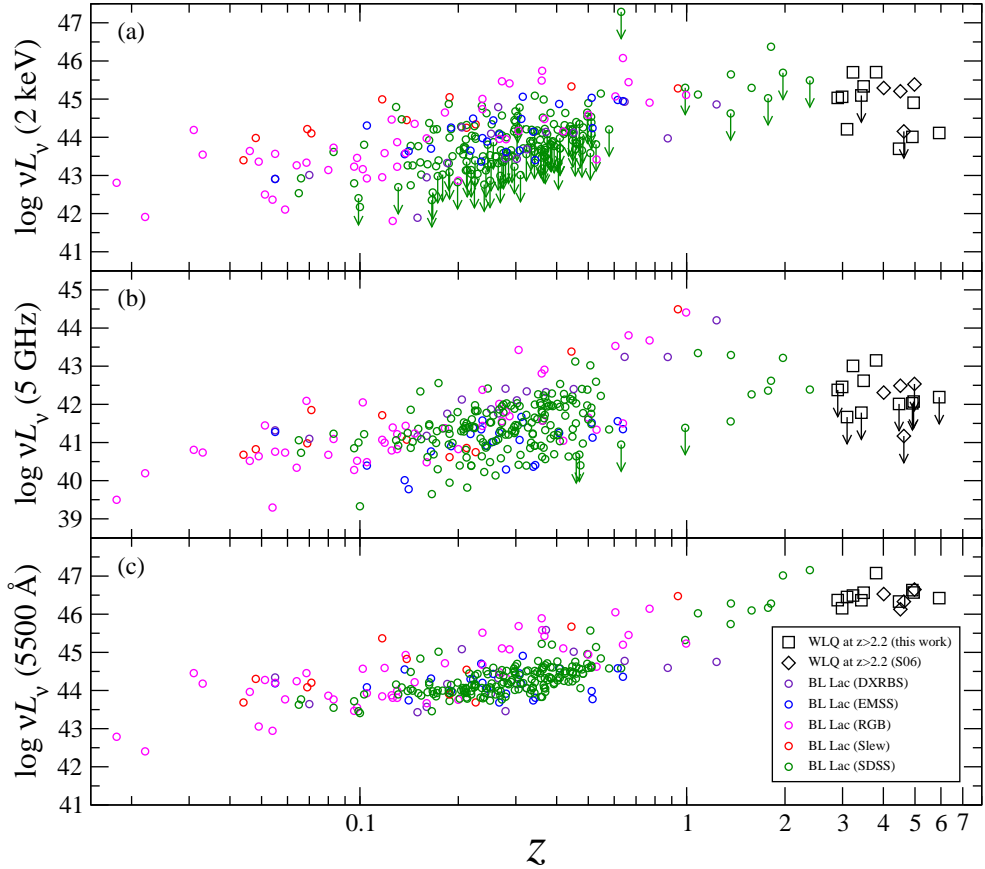}
 \caption{Monochromatic luminosities (in units of ergs\,s$^{-1}$) in
   the \xray\ (a), radio (b), and optical (c) bands versus redshift
   for BL~Lacs and WLQs. Squares mark $z>2.2$ WLQs from this work, and
   diamonds mark $z>2.2$ WLQs from S06. Violet, blue, magenta, and red
   circles mark BL~Lacs from the DXRBS (Perlman \et 1998; Landt \et
   2001), EMSS (e.g., Rector \et 2000), RGB (e.g., Laurent-Muehleisen
   \et 1999), and the {\sl Einstein} Slew (e.g., Perlman \et 1996)
   surveys, respectively. Green circles mark higher confidence SDSS
   BL~Lac candidates with reliable redshifts from the SDSS (Collinge
   \et 2005; Plotkin \et 2008). Upper limits are marked with
   down-pointing arrows.}
\label{L_z}
\end{figure*}

\subsection{Comparing WLQs with BL~Lacs}
\label{BLLac}

For the purpose of comparing our WLQs to BL~Lac objects, we have
compiled a dataset of \xray, optical, and radio luminosities for a
sample of 279 BL~Lacs from a variety of surveys, covering most of the
current BL~Lac selection methods (see, e.g., Perlman \et 2001). We
selected only BL~Lacs with reliable redshifts from the {\sl Einstein}
Slew survey (11 sources; e.g., Perlman \et 1996), the {\sl Einstein}
Medium Sensitivity Survey (EMSS; 24 sources; e.g., Rector \et 2000),
the {\sl ROSAT} All-Sky Survey-Green Bank (RGB) catalog (47 sources;
e.g., Laurent-Muehleisen \et 1999), the Deep \xray\ Radio Blazar
Survey (DXRBS; 16 sources; Perlman \et 1998; Landt \et 2001), and the
SDSS (181 sources; Collinge \et 2005; Plotkin \et 2008). Most BL\,Lacs
with reliable redshifts from the 1\,Jy survey (e.g., Stickel \et
1991), i.e., radio-selected BL~Lacs, overlap with sources detected in
the above surveys and are included in our sample.\,Basic \xray,
optical, and radio properties of our~BL Lac sample are given in
Table\,\ref{BLLac_sample} (placed\,at\,the\,end\,of\,the\,paper).

\setcounter{table}{5}

All fluxes and $K$-corrections for our BL\,Lac sample were computed
assuming the spectral slopes $\alpha_{\rm x}=-1$, $\alpha_{\rm
  o}=-0.5$, and $\alpha_{\rm r}=-0.5$ (where \hbox{$f_{\nu}\propto
  \nu^{\alpha}$}) for the \xray, optical--UV, and radio bands,
respectively; the same as those assumed for our WLQs. These spectral
slopes are consistent with the observed values in the majority of
quasars, although in the optical and radio bands they are somewhat
different from the slopes observed in typical BL\,Lacs (e.g., Sambruna
et 1996; Rector \et 2000; Reeves \& Turner 2000; Vanden Berk \et 2001;
Collinge \et 2005). We repeated the optical and radio flux and
$K$-correction calculations for the BL\,Lacs, this time assuming
typical observed BL\,Lac slopes (i.e., $\alpha_{\rm r}=0.27$; e.g.,
Stickel \et 1991; $\alpha_{\rm o}=-1.5$; e.g., Collinge \et 2005), and
found that the differences in optical and radio fluxes between the two
sets of calculations were small (since the majority of these BL\,Lacs
lie at $z\ltsim0.3$).\,In\,any\,case,\,the differences in optical and
radio fluxes between the calculations using different slopes are well
within the uncertainties stemming from measurement
errors,\,BL\,Lac\,variability,\,and\,beaming\,(note\,that, for
consistent comparisons with the observed\,properties of\,the\,WLQs, we
make no attempt to correct the fluxes or luminosities of the BL\,Lacs
for the effects of relativistic beaming).

In Figure~\ref{L_z} we present \xray, optical, and radio monochromatic
luminosities versus redshift for our BL~Lac comparison sample and our
extended sample of 15 WLQs. In spite of the large difference in
redshift between the majority of BL~Lacs ($z\ltsim0.3$) and WLQs
($z\gtsim2.2$), the WLQs do not have luminosities that are so high
that they are disjoint from those of the BL~Lacs; a few BL~Lacs even
reach the optical luminosities of WLQs. The fact that no BL~Lacs have
been found in the redshift range of our WLQs is not yet understood
(e.g., Stocke \& Perrenod 1981; Stocke 2001; Padovani \et 2007).

\begin{figure*}
\epsscale{1.2}
 \plotone{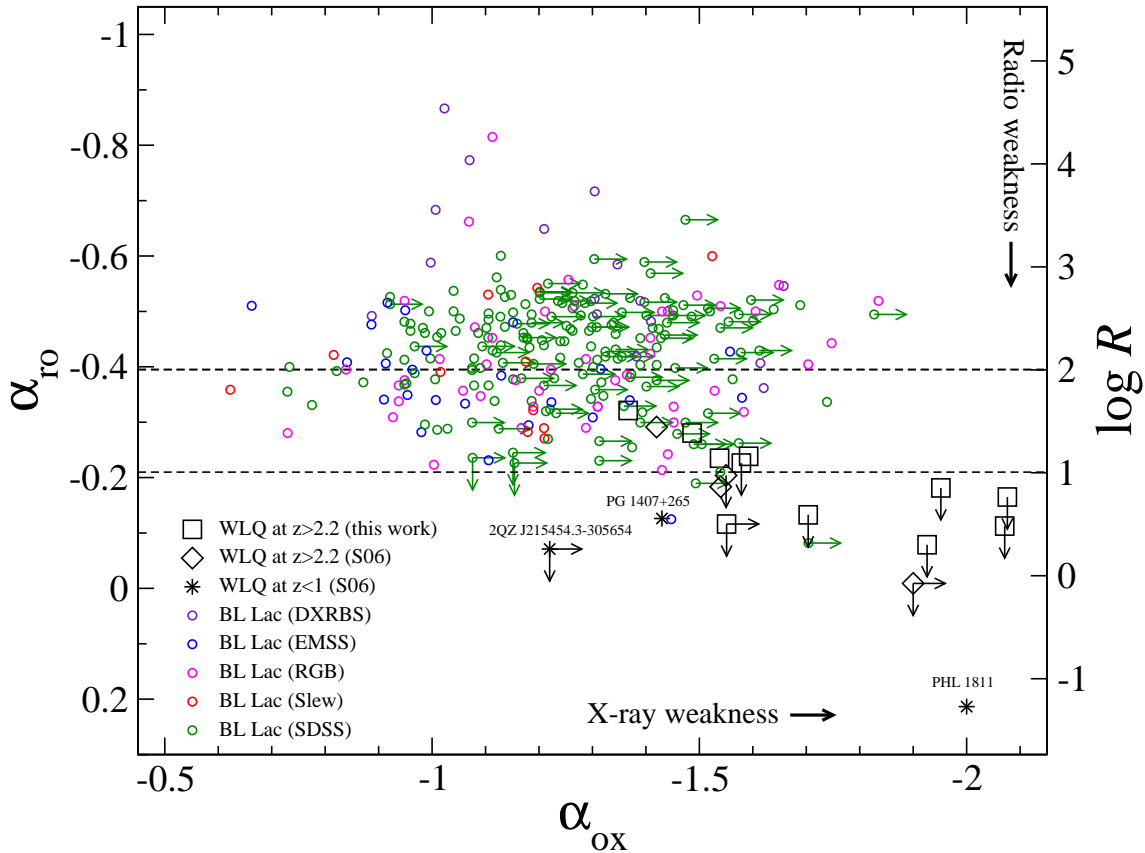}
 \caption{The \aox--\aro\ diagram for BL~Lacs and WLQs. Symbols are
   similar to those in Fig.~\ref{L_z}. WLQs at $z<1$ from S06 are
   marked with asterisks. Right (down) pointing arrows mark \aox\
   (\aro) upper limits. All sources that lie above (below) the upper
   (lower) dashed line are radio loud (quiet) with $R>100$ ($R<10$).}
\label{a_ro_a_ox}
\end{figure*}

Are WLQs simply the long-sought high-redshift BL~Lacs? In this
scenario, it is difficult to explain why there are essentially no
\xray- and/or radio-bright WLQs in Figure~\ref{a_ro_a_ox}, where we
have plotted \aox\ and \aro\ for the BL~Lacs and our WLQs in a
`color-color' diagram (the definitions of \aox\ and \aro\ in
\S\,\ref{basic} are identical to the definitions used in S06, and note
that the similar Fig.~8 of S06 included many flat-spectrum radio
sources and sources with unreliable redshifts). In other words, if the
SDSS high-redshift WLQs are indeed the high-redshift tail of the
BL~Lac population with \aox\ and \aro\ distributions similar to their
low-redshift counterparts, then a much larger, `parent' population of
\xray- and radio-bright lineless sources at high redshift is still
missing. Unless these are too optically faint (see below), the SDSS,
for example, should have naturally discovered many such sources had
they existed (see, e.g., Collinge \et 2005; Anderson \et 2007; Plotkin
\et 2008). Alternatively, if most BL~Lacs at low redshift were \xray\
and radio weak, and thus were largely missed by \xray\ and radio
surveys, the SDSS should have discovered many members of this
population as well (e.g., Collinge \et 2005).

To test whether \xray- and radio-bright BL~Lacs are simply too faint
and thus miss detection at high redshift, we looked for correlations
between luminosity and \aox\ and \aro\ among our BL~Lac sample. We
found that \aox\ and \aro\ are not significantly correlated with
optical luminosity. On the other hand, \aox\ (\aro) is significantly
correlated with \xray\ (radio) luminosity in the sense that, sources
that are more \xray\ (radio) luminous are also \xray\ (radio) brighter
relative to their optical luminosities. These results suggest that we
are not missing a putative, large parent population of BL~Lacs at high
redshift purely due to a luminosity effect. An additional difficulty
in associating WLQs with BL~Lacs arises from the fact that, while
there appears to be a trend of a positive \xray-radio dependence among
WLQs, there is no corresponding trend between the \xray\ and radio
emission of BL~Lacs (i.e., for a given \aox\ value, a BL~Lac can have
a wide range of \aro\ values and vice versa; Fig.~\ref{a_ro_a_ox}; see
also Plotkin \et 2008).

\begin{figure}
 \plotone{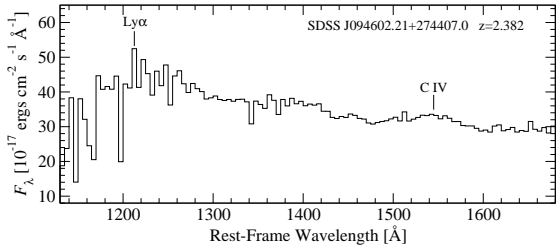}
 \caption{The SDSS spectrum of the BL~Lac candidate
   SDSS~J094602.21$+$274407.0 from Plotkin \et (2008), which may also
   be classified as a WLQ. The spectrum has been resampled in bins of
   5\,\AA\ (as in Fig.~\ref{spectra}). The locations of two prominent
   UV emission lines have been marked.}
\label{sdssj0946}
\end{figure}

Are WLQs related to the rare population of radio-weak BL~Lacs (see,
e.g., Collinge \et 2005; Anderson \et 2007)? Four BL~Lacs from our
comparison sample, SDSS~J095125.90$+$504559.7 at $z=1.363$,
SDSS~J094602.21$+$274407.0 at $z=2.382$, SDSS~J145326.52$+$545322.4 at
$z=0.100$ (Plotkin \et 2008), and MS~2306.1$-$2236 at $z=0.137$
(Rector \et 2000) have \aro$>-0.2$ and satisfy the radio-weak BL~Lac
(as well as radio-quiet AGN) classification; see also Table~5 of
Collinge \et (2005) for additional possible candidates from this
group. Based on its redshift and UV properties,
SDSS~J094602.21$+$274407.0 may be classified as a high-redshift WLQ
(it is the radio-weakest source with an \aox\ upper limit from the
SDSS BL~Lac sample in Fig.~\ref{a_ro_a_ox}); we plot its SDSS spectrum
in Figure~\ref{sdssj0946}. Further testing the association of WLQs
with radio-weak BL~Lacs requires broader spectroscopic coverage of
both classes. For example, near-IR spectra of WLQs should be acquired
to search for low-ionization emission lines in the rest-frame
UV--optical band, and UV spectra of low-redshift, radio-weak BL~Lacs
should be obtained to measure the strengths of high-ionization
emission lines. However, such observations will not resolve the
missing parent population of `typical' (i.e., \xray- and radio-bright)
BL~Lacs at high redshift, which remains an outstanding problem for any
model attempting to unify WLQs and BL Lacs.

\subsection{Are WLQs Extreme Quasars?}
\label{extreme}

The second scenario attempting to explain the missing UV lines in WLQs
suggests that these sources are quasars hosting abnormal BELRs or
quasars with extreme properties (\S\,\ref{introduction}). In this
case, one does not expect to find many radio-bright (i.e., radio-loud)
WLQs, since the majority of quasars at all luminosities and redshifts
are known to be radio quiet (e.g., Jiang \et 2007). Hence, unless
emission-line strength was correlated with radio properties, there is
no problem of a missing parent population (see \S\,\ref{BLLac}). In
addition, the WLQs in our sample show blue UV continua with an average
slope \hbox{$\alpha_{\rm UV}=-0.8$}, consistent with the slopes
observed for high-redshift quasars (e.g., Schneider \et 2001) as well
as with the slopes of a larger sample of high-redshift WLQs
(Diamond-Stanic \et 2009). This apparently steeper slope (compared
with \hbox{$\alpha_{\rm UV}=-0.5$} observed for low-redshift quasars)
is essentially due to the restricted wavelength range over which this
slope can be measured in the optical spectra of high-redshift sources
(i.e., given a sufficiently broad wavelength range, the UV continuum
slope is \hbox{$\alpha_{\rm UV}=-0.5$}, and it does not evolve with
redshift; see, e.g., Schneider \et 2005). BL~Lacs, on the other hand,
are considerably redder in this band (with \hbox{$\alpha_{\rm
    o}\approx -1.5$}; e.g., Collinge \et 2005). The lack of red UV
spectra in WLQs may also be reflected by the upper limit we place on
the mean column density of neutral absorption in these sources
(\S\,\ref{sec_joint_fit}; Table~\ref{joint_fit_par}). This is broadly
consistent with the idea that the weakness of the emission lines is
not a consequence of dust absorption, even though the upper limit on
\nh\ does not provide a tight constraint on allowed reddening in the
rest-frame UV.

Another indication that WLQs are related to typical quasars lies in
their \aox\ distribution (with a spread of $\sim0.6$ for the
radio-quiet sources; see Fig.~\ref{a_ro_a_ox}). We tested whether this
\aox\ distribution is consistent with the optically selected sample of
Gibson \et (2008), that includes SDSS quasars with \chandra\
detections. We performed a KS test and found that the \aox\
distributions were consistent between 139 radio-quiet, non-BAL quasars
from sample~B of Gibson \et (2008) and our seven radio-quiet WLQs. We
found that the null hypothesis, that the two samples were drawn from
the same parent distribution, cannot be rejected with a confidence
level of 97.3\%. After scaling out the dependence of \aox\ on optical
luminosity (\S\,\ref{results}), we also performed a KS test on the two
\daox\ distributions of these samples and found that the null
hypothesis cannot be rejected with a confidence level of 94.8\%.

As noted in S06, at least two sources at low redshift, PG~1407$+$265
at $z=0.94$ (McDowell \et 1995) and PHL~1811 at $z=0.19$ (Leighly \et
2007b), lack high-ionization emission lines in their UV spectra
(although some low-ionization lines are detected in both sources). An
additional source, 2QZ~J215454.3$-$305654 at $z=0.494$ (Londish \et
2004), may also be related to these two, as it shows only an
[\ion{O}{3}]\,$\lambda5007$ emission line; however, since UV
spectroscopy has not yet been performed, it is unknown whether it
displays higher ionization lines. These three sources, which also
appear different from BL~Lacs, may perhaps be the low-redshift analogs
of the $z\gtsim2.2$ WLQs; they are plotted separately in
Figure~\ref{a_ro_a_ox}. The fact that all our WLQs lie at high
redshifts is clearly a selection effect due to the requirement of a
Lyman break detection. We therefore expect that WLQs should exist also
at lower redshifts.

\begin{figure}
\epsscale{1.25}
 \plotone{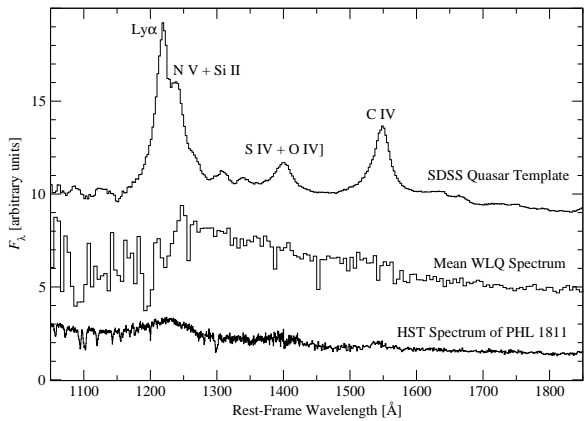}
 \caption{A comparison between the mean UV spectrum of $\sim2000$ SDSS
   quasars ({\it top}; Vanden Berk \et 2001), the mean UV spectrum of
   ten of our WLQs with SDSS spectra ({\it middle}; see also
   Fig.~\ref{spectra}), and the {\sl HST/STIS} UV spectrum of PHL~1811
   ({\it bottom}; Leighly \et 2007b). All spectra are scaled
   arbitrarily in flux for clarity. Prominent emission lines are
   marked on the mean SDSS spectrum.}
\label{templates}
\end{figure}

Figure~\ref{templates} shows a comparison between the UV spectrum of
one of these low-redshift WLQs, PHL~1811, and the mean UV spectrum of
ten of our high-redshift WLQs (i.e., excluding the Keck spectrum of
SDSS~J1335$+$3533) alongside the mean UV spectrum of `typical' SDSS
quasars. Overall, the mean spectrum of our high-redshift WLQs appears
similar to the spectrum of PHL~1811, and the high-ionization emission
lines in both spectra are significantly weaker compared to the mean
SDSS spectrum. The Ly$\alpha$+\ion{N}{5} complex in PHL~1811 has
EW$=24$\,\AA, which is somewhat larger than our definition of WLQs
(EW$<10$\,\AA), but the \ion{C}{4} emission line in that source is
exceptionally weak with EW$=6.6$\,\AA\ (Leighly \et 2007b). A
comprehensive comparison between our high-redshift WLQs and their
putative low-redshift counterparts requires rest-frame optical
spectroscopy of our sources to search for low-ionization emission
lines.

Weak emission lines in AGNs are typically associated with high
luminosities in what has been termed the `Baldwin effect', i.e., the
anticorrelation between emission-line EW and quasar luminosity (where
the anticorrelation is stronger and steeper for emission lines with
higher ionization potentials; e.g., Baldwin 1977; Dietrich \et
2002). In this context, WLQs might have been expected to dominate the
high-luminosity end of the quasar luminosity function, but this is
certainly not the case (e.g., Just \et 2007 studied the properties of
the 32 most luminous quasars in SDSS~DR3; this sample does not include
a single WLQ).

A different interpretation of the Baldwin effect has been suggested by
Baskin \& Laor (2004), who have shown that EW(\ion{C}{4}) depends more
strongly on the normalized accretion rate (i.e., \lledd, where $L$ is
the AGN bolometric luminosity) than on $L_{\rm UV}$ in a sample of 81
low--moderate luminosity AGNs (i.e., with
\hbox{$L\sim$10$^{43}$--10$^{46}$}\,ergs~s$^{-1}$) from Boroson \&
Green (1992).  Using virial black-hole masses measured from the SDSS
spectra of $\sim60000$ quasars, Shen \et (2008) have shown that
\lledd\ exhibits a lognormal distribution (clearly, WLQs could not
have been included in their analysis due to their lack of emission
lines). Can this be related to the result of Diamond-Stanic \et (2009)
who found that EW(Ly$\alpha +$\ion{N}{5}) values in SDSS quasars also
exhibit a lognormal distribution (\S~\ref{introduction})? In this
scenario, WLQs may constitute a significant deviation from the \lledd\
distribution at the high-\lledd\ end, in analogy with their
significant deviation from the EW distribution at the low-EW end.
Testing this idea and the hypothesis that emission-line EW is
correlated primarily with \lledd\ across the entire AGN luminosity
range requires extending the Baskin \& Laor (2004) analysis to much
higher luminosities, such as those of our WLQs; this is beyond the
scope of this work. Near-IR spectroscopy of the \ion{Mg}{2} and
H$\beta$ spectral regions in WLQs is required to determine \lledd\
values in these sources directly, provided these lines are
sufficiently strong to measure (e.g., Shemmer \et 2004). A high
accretion rate has also been proposed for at least one of the
low-redshift WLQs, PHL~1811, to explain the absence of the
high-ionization emission lines (Leighly \et 2007b).

If WLQs are AGNs that constitute the high-end tail of the \lledd\
distribution, then they may also be expected to exhibit steep
hard-\xray\ continua (see, e.g., Brandt \et 1997; Shemmer \et
2008). For example, high-quality \xmm\ spectra of two of the
low-redshift WLQs, PG~1407$+$265 and PHL~1811, indeed show relatively
steep \xray\ continua at rest-frame energies $\gtsim2$\,keV, with
photon indices of $\Gamma=2.19\pm0.07$ (Piconcelli \et 2005) and
$\Gamma=2.28^{+0.12}_{-0.11}$ (Leighly \et 2007a),
respectively. High-quality \xray\ spectra of high-redshift WLQs have
not yet been obtained to test whether they show similar \xray\
spectral properties. The \xray\ brightest source in our sample,
SDSS~J1231$+$0138, had only 110 counts, but it does not exhibit an
unusually steep hard-\xray\ power-law spectrum
($\Gamma=1.76^{+0.35}_{-0.34}$; \S\,\ref{sdssj1231}). However, this
WLQ is also the radio-brightest source in our sample (involving
perhaps jet-related emission contributing to its \xray\ spectrum). We
had already pointed out that our joint spectral fitting of a sample of
WLQs does not provide any evidence or hint of a steep hard-\xray\
spectrum (\S\,\ref{sec_joint_fit}; Table~\ref{joint_fit_par}).
Moreover, if we consider WLQs to have \lledd\gtsim1 (i.e., at the
extreme high end of the Shen \et 2008 \lledd\ distribution), then
following the $\Gamma$-\lledd\ correlation of Shemmer \et (2008) these
sources should have hard-\xray\ slopes with $\Gamma$\gtsim2.5. Such
steep slopes are rejected with $>$99.9\% confidence from our joint
spectral fits. Interestingly, however, Table~\ref{counts} may hint
that at least two WLQs, SDSS~J0317$-$0758 and SDSS~J0928$+$1848 (with
insufficient counts for a proper spectral analysis), may have steep
hard \xray\ spectra, based on the effective photon indices obtained
from their band ratios. Future \xmm\ spectroscopic observations of
these as well as a statistically representative sample of WLQs are
required for a robust test of this idea. Such observations may also
provide better constraints on putative Compton reflections in these
sources.

\section{Summary}
\label{summary}

We present new \chandra\ observations of nine high-redshift quasars
with weak or undetectable high-ionization emission lines in their UV
spectra (WLQs). Adding archival \xray\ observations of six additional
sources of this class has enabled us to place the strongest
constraints yet on the \xray\ properties of 15 sources of this rare
class of AGNs. All but two of these 15 sources are \xray\ detected,
and for the strongest \xray\ (and radio) source we have determined a
power-law photon index of $\Gamma=1.76^{+0.35}_{-0.34}$ in the
rest-frame $\sim2-35$\,keV spectral band. A joint spectral fit of the
\xray\ spectrum of 11 of these sources gives an upper limit on the
mean intrinsic neutral absorption column density of \nh$\ltsim5\times
10^{22}$\,cm$^{-2}$; the corresponding mean power-law photon index is
$\Gamma=1.79^{+0.17}_{-0.16}$. These results provide further support
to the idea that the weakness of the UV lines is not a consequence of
dust absorption, and they show that the mean hard-\xray\ spectral
slope of WLQs is consistent with the spectral slopes observed in
typical radio-quiet quasars.

Our new \xray\ information on high-redshift WLQs has brought us one
step closer to distinguishing between two competing models for their
nature. Although a boosted-continuum interpretation for the missing
emission lines cannot be ruled out based on our current data, we find
that WLQs occupy an almost distinct region of the \xray-radio
parameter space from BL~Lacs. WLQs are mostly \xray\ and radio weak
compared to lower-redshift BL~Lacs. The strongest argument against the
association of WLQs with BL~Lacs is the lack of a parent population of
\xray- and radio-bright sources at high redshift. We discuss
additional observations that are required to test whether WLQs are
quasars with extreme properties. A key goal is to check whether WLQs
are quasars with extremely high accretion rates; our current
constraints on the mean $\Gamma$ for these sources do not support this
idea. Such efforts should provide a better understanding not only of
WLQs and the BL~Lac phenomena, but of the AGN population as a whole.

\acknowledgments

We gratefully acknowledge the financial support of \chandra\ \xray\
Center grant \hbox{G07-8101X}, NASA LTSA grant \hbox{NAG5-13035}
(O.~S., W.~N.~B., D.~P.~S.), and NSF grant AST-0702766 (M.~A.~S.). We
thank Annalisa Celotti, Abe Falcone, and Eric Perlman for fruitful
discussions. We also thank an anonymous referee for a thoughtful
report that assisted in improving the presentation of this work.
Funding for the SDSS and SDSS-II has been provided by the Alfred
P. Sloan Foundation, the Participating Institutions, the National
Science Foundation, the U.S. Department of Energy, the National
Aeronautics and Space Administration, the Japanese Monbukagakusho, the
Max Planck Society, and the Higher Education Funding Council for
England. The SDSS Web Site is http://www.sdss.org/. The SDSS is
managed by the Astrophysical Research Consortium for the Participating
Institutions. The Participating Institutions are the American Museum
of Natural History, Astrophysical Institute Potsdam, University of
Basel, University of Cambridge, Case Western Reserve University,
University of Chicago, Drexel University, Fermilab, the Institute for
Advanced Study, the Japan Participation Group, Johns Hopkins
University, the Joint Institute for Nuclear Astrophysics, the Kavli
Institute for Particle Astrophysics and Cosmology, the Korean
Scientist Group, the Chinese Academy of Sciences (LAMOST), Los Alamos
National Laboratory, the Max-Planck-Institute for Astronomy (MPIA),
the Max-Planck-Institute for Astrophysics (MPA), New Mexico State
University, Ohio State University, University of Pittsburgh,
University of Portsmouth, Princeton University, the United States
Naval Observatory, and the University of Washington. This research has
made use of the NASA/IPAC Extragalactic Database (NED) which is
operated by the Jet Propulsion Laboratory, California Institute of
Technology, under contract with the National Aeronautics and Space
Administration.

\setcounter{table}{2}

\clearpage

\begin{landscape}
\begin{deluxetable}{lcccrcrrrrrrrr}
\tablecolumns{14}
\tablewidth{0pt}
\tablecaption{X-ray, Optical, and Radio Properties}
\tablehead{
\colhead{} &
\colhead{} &
\colhead{} &
\colhead{} &
\colhead{} &
\colhead{$\log(\nu L_\nu)$} &
\colhead{Count} & 
\colhead{} &
\colhead{} &
\colhead{$\log (\nu{L_\nu})$} &
\colhead{$\log L$} &
\colhead{} &
\colhead{} &
\colhead{} \\
\colhead{Object (SDSS~J)} &
\colhead{$N_{\rm H}$\tablenotemark{a}} &
\colhead{$AB_{1450}$} &
\colhead{$M_B$} &
\colhead{$f_{2500~{\rm \AA}}$\tablenotemark{b}} &
\colhead{2500~\AA} &
\colhead{Rate\tablenotemark{c}} &
\colhead{$f_{\rm x}$\tablenotemark{d}} &
\colhead{$f_{2~\rm keV}$\tablenotemark{e}} &
\colhead{${2~\rm keV}$} &
\colhead{2--10~keV} &
\colhead{$\alpha_{\rm ox}$} &
\colhead{\daox~$(\sigma)$\tablenotemark{f}} &
\colhead{$\alpha_{\rm ro}$\tablenotemark{g}} \\
\colhead{(1)} &
\colhead{(2)} &
\colhead{(3)} &
\colhead{(4)} &
\colhead{(5)} &
\colhead{(6)} &
\colhead{(7)} &
\colhead{(8)} &
\colhead{(9)} &
\colhead{(10)} &
\colhead{(11)} &
\colhead{(12)} &
\colhead{(13)} &
\colhead{(14)}
}
\startdata
031712.23$-$075850.3 & 5.14 & 18.7 & $-$27.1 & 15.8 &
46.5 & {\phn}5.40$^{+1.35}_{-1.10}$ & {\phn}22.1$^{+5.5}_{-4.5}$ &
12.19 & {\phn}45.0 & {\phn}45.2 & $-1.58$ & $+0.12 \ (0.8)$ &
$>-0.23$\tablenotemark{l} \\
092832.90$+$184824.4 & 3.83 & 17.4 & $-$28.9 & 52.1 &
47.2 & {\phn}13.13$^{+2.08}_{-1.81}$ & {\phn}51.7$^{+8.2}_{-7.1}$ &
37.03 & {\phn}45.7 & {\phn}45.9 & $-1.59$ & $+0.21 \ (1.6)$ & $-0.24$ \\
101204.04$+$531331.8\tablenotemark{h} & 0.78 & 19.3 &
$-$26.6 & 9.2 & 46.3 & {\phn}3.28$^{+4.85}_{-1.86}$\tablenotemark{k} &
{\phn}20.8$^{+30.7}_{-11.8}$ & 12.37 & {\phn}45.1 & {\phn}45.3 &
$-1.49$ & $+0.18 \ (1.3)$ & $-0.28$ \\
114153.34$+$021924.3 & 2.30 & 18.5 & $-$27.6 & 18.4 &
46.7 & {\phn}7.18$^{+1.61}_{-1.33}$ & {\phn}27.1$^{+6.1}_{-5.0}$ &
18.07 & {\phn}45.3 & {\phn}45.5 & $-1.54$ & $+0.19 \ (1.3)$ & $-0.24$ \\
121221.56$+$534128.0 & 1.71 & 18.6 & $-$27.3 & 17.2 &
46.6 & {\phn}0.73$^{+0.71}_{-0.39}$ & {\phn}2.7$^{+2.6}_{-1.5}$ & 1.65
& {\phn}44.2 & {\phn}44.4 & $-1.93$ & $-0.21 \ (1.5)$ & $>-0.08$ \\
123132.38$+$013814.0 & 1.81 & 18.6 & $-$27.4 & 17.5 &
46.7 & {\phn}20.44$^{+2.49}_{-2.23}$ & {\phn}76.0$^{+9.3}_{-8.3}$ &
47.94 & {\phn}45.7 & {\phn}45.9 & $-1.37$
& $+0.35 \ (2.4)$ & $-0.32$ \\
123743.08$+$630144.9\tablenotemark{h} & 1.66 & 19.0 &
$-$27.1 & 12.0 & 46.5 & {\phn}$<3.30$\tablenotemark{k} & {\phn}$<16.6$
& $<10.88$ &
{\phn}$<45.1$ & {\phn}$<45.3$ & $<-1.55$ & $<+0.15 \ (<1.0)$ & $>-0.12$ \\
130216.13$+$003032.1\tablenotemark{i} & 1.57 & 19.5 &
$-$27.0 & 7.3 & 46.5 & {\phn}0.09$^{+0.12}_{-0.06}$ &
{\phn}0.3$^{+0.5}_{-0.2}$ & 0.28 & 43.7 & 43.9 & $-2.08$ & $-0.38 \
(2.6)$ & $>-0.16$ \\
133422.63$+$475033.6 & 1.63 & 19.1 & $-$27.6 & 10.7 &
46.7 & {\phn}1.19$^{+0.41}_{-0.32}$ & {\phn}4.4$^{+1.5}_{-1.2}$ & 3.90
& {\phn}44.9 & {\phn}45.1 & $-1.70$ &
$+0.02 \ (0.2)$ & $>-0.13$ \\
133550.81$+$353315.8 & 0.92 & {\phn}19.8\tablenotemark{j} & $-$27.3 &
5.8 & 46.6 & {\phn}0.13$^{+0.13}_{-0.07}$ & {\phn}0.5$^{+0.5}_{-0.3}$
& 0.48 & {\phn}44.1 & {\phn}44.3 & $-1.95$ & $-0.24 \ (1.7)$ &
$>-0.18$ \\
142103.83$+$343332.0 & 1.16 & 19.0 & $-$27.8 & 12.5 & 46.8 &
{\phn}0.16$^{+0.21}_{-0.10}$ & {\phn}0.6$^{+0.8}_{-0.4}$ & 0.50 &
{\phn}44.0 & {\phn}44.2 & $-2.07$ & $-0.33 \ (2.3)$ & $>-0.11$
\enddata
\tablenotetext{a}{Neutral Galactic absorption column density in units
  of $10^{20}$~cm$^{-2}$ taken from Dickey \& Lockman (1990).}
\tablenotetext{b}{Flux density at rest-frame 2500~\AA\ in units of
  10$^{-28}$ erg cm$^{-2}$ s$^{-1}$ Hz$^{-1}$.}
\tablenotetext{c}{Observed count rate computed in the
  \hbox{0.5--2~keV} band in units of $10^{-3}$ counts s$^{-1}$.}
\tablenotetext{d}{Galactic absorption-corrected flux in the observed
  \hbox{0.5--2~keV} band in units of $10^{-15}$ erg cm$^{-2}$
  s$^{-1}$.}
\tablenotetext{e}{Flux density at rest-frame 2~keV in units of
  $10^{-32}$ erg cm$^{-2}$ s$^{-1}$ Hz$^{-1}$.}
\tablenotetext{f}{The difference between measured and predicted \aox\
  ($\Delta$\aox), and the significance of that difference ($\sigma$),
  based on the Just \et (2007) \aox--$L_{\nu}(2500~\mbox{\AA})$
  relation.}
\tablenotetext{g}{Unless otherwise noted, radio-to-optical flux ratios
  ($\alpha_{\rm ro}$; see \S\,\ref{basic} for the conversion of these
  values to the radio-loudness parameter $R$) involve radio flux
  densities at an observed-frame frequency of 1.4~GHz taken from the
  FIRST survey (Becker \et 1995); upper limits on $\alpha_{\rm ro}$
  are calculated from the 3~$\sigma$ FIRST detection threshold at the
  source position.}
\tablenotetext{h}{\xray\ data obtained from serendipitous \xmm\
  observations.}
\tablenotetext{i}{The \xray\ count rate, fluxes, luminosities, and
  \aox\ were obtained from the merged event file of two \chandra\
  exposures (see \S~\ref{observations}).}
\tablenotetext{j}{Obtained from the Keck spectrum of Fan \et (2006).}
\tablenotetext{k}{Count rate corrected for the exposure map.}
\tablenotetext{l}{Flux density at an observed-frame frequency of
  1.4~GHz taken from the NVSS survey (Condon \et 1998); upper limits
  are calculated from the NVSS detection threshold of 2.5~mJy.}
\label{properties}
\end{deluxetable}
\clearpage

\end{landscape}

\setcounter{table}{4}

\clearpage

\begin{landscape}
\begin{deluxetable}{lrrcrrrrrc}
\tablecolumns{10}
\tablewidth{0pt}
\tablecaption{X-ray, Optical, and Radio Properties of the BL~Lac Sample}
\tablehead{
\colhead{} &
\colhead{RA} &
\colhead{DEC} &
\colhead{} &
\colhead{$\log [\nu L_{\nu} ({2~\rm keV})]$} &
\colhead{$\log [\nu L_{\nu} (5500~\mbox{\AA})]$} &
\colhead{$\log [\nu L_{\nu} (5~{\rm GHz})]$} &
\colhead{} &
\colhead{} &
\colhead{} \\
\colhead{BL~Lac} &
\colhead{(J2000)} &
\colhead{(J2000)} &
\colhead{$z$} &
\colhead{(ergs~s$^{-1}$)} &
\colhead{(ergs~s$^{-1}$)} &
\colhead{(ergs~s$^{-1}$)} &
\colhead{$\alpha_{\rm ox}$} &
\colhead{$\alpha_{\rm ro}$} &
\colhead{Reference}
}
\startdata
SDSS~J000157.23$-$103117.3 & 0.48850 & $-10.52148$ & 0.252 & $\leq43.30$ &
$44.05$ & $41.53$ & $\leq-1.35$ & $-0.50$ & 1 \\
RGB~J0007$+$472 & 2.00000 & $47.20222$ & 0.280 & $43.95$ & $44.33$ &
$41.81$ & $-1.21$ & $-0.50$ & 2 \\
SDSS~J001736.91$+$145101.9 & 4.40378 & $14.85053$ & 0.303 & $\leq44.09$ &
$44.70$ & $42.03$ & $\leq-1.30$ & $-0.47$ & 3 \\
SDSS~J002200.95$+$000657.9 & 5.50396 & $0.11610$ & 0.306 & $44.24$ &
$44.03$ & $40.42$ & $-0.99$ & $-0.30$ & 1 \\
WGA~J0032.5$-$2849 & 8.13792 & $-28.82222$ & 0.324 & $43.68$ & $44.06$ &
$42.34$ & $-1.21$ & $-0.65$ & 4 \\
WGA~J0043.3$-$2638 & 10.84417 & $-26.65139$ & 0.451 & $44.12$ & $45.02$
& $42.40$ & $-1.41$ & $-0.48$ & 4 \\
SDSS~J005620.07$-$093629.7 & 14.08366 & $-9.60826$ & 0.103 & $43.80$ &
$43.90$ & $41.22$ & $-1.10$ & $-0.47$ & 1 \\
WGA~J0100.1$-$3337 & 15.03917 & $-33.62556$ & 0.875 & $43.97$ & $44.59$
& $43.24$ & $-1.30$ & $-0.72$ & 4 \\
SDSS~J010326.01$+$152624.8 & 15.85839 & $15.44021$ & 0.246 & $\leq43.81$ &
$44.43$ & $42.42$ & $\leq-1.30$ & $-0.59$ & 3 \\
RGB~J0110$+$418 & 17.52000 & $41.83083$ & 0.096 & $43.23$ & $43.47$ &
$40.28$ & $-1.16$ & $-0.38$ & 2
\enddata
\tablecomments{The full version of this table is available in the
  electronic edition online. Luminosities, \aox, and \aro\ were
  computed from the data given in the references (last column) using
  the $K$-corrections outlined in \S~\ref{BLLac}.}
\tablerefs{(1) Plotkin \et (2008); (2) Laurent-Muehleisen \et (1999);
  (3) Collinge \et (2005); (4) Perlman \et (1998).}
\label{BLLac_sample}
\end{deluxetable}

\clearpage

\end{landscape}

\end{document}